\documentclass[floats,floatfix,showpacs,amssymb,prd,twocolumn,superscriptaddress,nofootinbib,nolongbibliography,reprint]{revtex4-2}

\usepackage{amssymb,amsmath,verbatim,mathtools,needspace,enumitem,etoolbox,graphicx,physics,microtype,afterpage,xspace,tabularx,lmodern,multirow}
\usepackage{gensymb}
\usepackage[dvipsnames, usenames]{xcolor}
\definecolor{linkcolor}{rgb}{0.0,0.3,0.5}
\usepackage[unicode, colorlinks=true, linkcolor=linkcolor, citecolor=linkcolor, filecolor=linkcolor, urlcolor=linkcolor, linktocpage, breaklinks]{hyperref}
\usepackage[all]{hypcap}
\usepackage[T1]{fontenc}
\usepackage[utf8]{inputenc}
\usepackage[usenames,dvipsnames]{xcolor}
\hypersetup{colorlinks=true,citecolor=romared,linkcolor=romared,urlcolor=romared}

\definecolor{romared}{RGB}{142,0,28}

\newcommand{\be}{\begin{equation}}
\newcommand{\ee}{\end{equation}}

\def\be{\begin{equation}}
\def\ee{\end{equation}}
\newcommand{\beq}{\begin{eqnarray}}
\newcommand{\eeq}{\end{eqnarray}}

\usepackage{aas_macros}
\usepackage{makecell}
\usepackage{soul}
\usepackage[nolist,nohyperlinks]{acronym}

\begin{document}

\title{The fate of observers in circular motion}

\author{Antoine Leh\'ebel} 
\email{antoine.lehebel@tecnico.ulisboa.pt}
\affiliation{Centro de Astrof\'{\i}sica e Gravita\c c\~ao  - CENTRA,
Departamento de F\'{\i}sica, Instituto Superior T\'ecnico - IST,
Universidade de Lisboa - UL,
Av. Rovisco Pais 1, 1049-001 Lisboa, Portugal}

\author{Vitor Cardoso} 
\email{vitor.cardoso@tecnico.ulisboa.pt}
\affiliation{Centro de Astrof\'{\i}sica e Gravita\c c\~ao  - CENTRA,
Departamento de F\'{\i}sica, Instituto Superior T\'ecnico - IST,
Universidade de Lisboa - UL,
Av. Rovisco Pais 1, 1049-001 Lisboa, Portugal}
\affiliation{Niels Bohr International Academy, Niels Bohr Institute, Blegdamsvej 17, 2100 Copenhagen, Denmark}

\begin{abstract}
In Newtonian physics or in general relativity, energy dissipation causes observers moving along circular orbits to slowly spiral towards the source of the gravitational field. We show that the loss of energy has the same effect in \textit{any} theory of gravity respecting the weak equivalence principle, by exhibiting an intimate relation between the energy of a massive test particle and the stability of its orbit. Ultimately, massive particles either plunge, or are driven towards minima of the generalized Newtonian potential, where they become static.
In addition, we construct a toy metric which displays an unbound innermost stable circular orbit, allowing particles that reach this orbit to be expelled away.
\end{abstract}
\maketitle

\section*{Introduction}
The weak equivalence principle is a cornerstone of gravitational physics. It can be stated as follows: the trajectory of a freely falling test body is independent of its internal structure and composition. In a metric theory of gravity, the weak equivalence principle asserts that the trajectories of freely falling test bodies are geodesics of the metric. Very stringent experimental tests (see \cite{Will:2014kxa} for a detailed review) confirm that, to the best of our knowledge, nature respects this principle. 
Hence, geodesics have a privileged role in the investigation of spacetime. A test body placed near a massive object can follow a geodesic that always stays in the vicinity of this massive object, an orbit. If the spacetime possesses enough symmetry, one can define circular orbits, which are of particular interest for two reasons: they are simple, and they are often the endpoint of dynamical evolution, as some dissipative effects -- mostly radiative effects \cite{PhysRev.136.B1224,Cardoso:2020iji}, tidal heating \cite{1966AnAp...29..489Z} or collisions (e.g., \cite{2017Icar..282..195H}) as beautifully illustrated by the rings of Saturn -- tend to circularize the orbits in general. The spacetime can then be divided in regions where timelike circular orbits exist or not, and are stable or not. There might also exist, at specific radii, lightlike circular orbits, also known as light rings (or photon spheres in spherical symmetry).

Some results about the structure of these regions are already known in whole generality. Under certain conditions that we will present further, horizonless compact objects possess an even number of light rings, one of them at least being stable~\cite{Cardoso:2014sna,Cunha:2017qtt}. The classification and conditions for the relative positions of light rings and regions where unstable/stable timelike circular orbits exist or not were recently discussed~\cite{Delgado:2021jxd}. Several disconnected regions of stable timelike circular orbits might exist, giving rise to a distinction between standard and absolute innermost stable circular orbits (ISCOs). The standard ISCO is the inner edge of the region that is connected to spatial infinity by timelike circular orbits. 
Aside of this question, certain spacetimes possess an unusual property: orbits where a particle can remain immobile indefinitely, also known as static rings~\cite{Collodel:2017end,Teodoro:2020kok,Collodel:2021jwi,Teodoro:2021ezj}.

In this paper, we aim at answering some simple latent questions, in the widest possible framework. First, in Newton's theory, or for simple solutions of Einstein's general relativity, circular orbits tend to shrink because of dissipative effects\footnote{See also the discussion in the concluding remarks.}. To what extent does this statement depend on the theory under consideration? Can it depend on the field content of the theory? Can there exist convoluted spacetimes, even in general relativity, where this intuition fails? What role do the symmetries of the spacetime play in the validity of this statement? 

Second, for solutions that possess an ISCO, the test body usually falls towards the center once it has reached the ISCO. It has not much choice in doing so: it cannot follow a stable circular orbit anymore, and its energy is usually such that it is gravitationally bound, hence it cannot escape. Is it possible to conceive a spacetime with an ISCO where test bodies are gravitationally unbound? If so, a test body which spiraled down to the ISCO could a priori escape the attraction of the massive body, and be expelled away.

To answer these questions, we first introduce the framework of circular geodesics in the context of stationary, axisymmetric and circular geometries. We then establish a crucial relation between the radial variation of the energy, the stability of the orbit and the generalized Newton potential. We further simplify this relation in the particular case of spherically symmetric spacetimes. Finally, we propose a toy metric which  features an unbound ISCO.

\section*{Framework}

We shall consider an (asymptotically) stationary and axisymmetric spacetime. Let us label the associated Killing vectors $\xi^\mu$ and $\psi^\mu$ respectively. We further assume that the spacetime is circular, namely that 
$\xi^\mu R_\mu^{\;[\nu}\xi^\rho\psi^{\sigma]}=\psi^\mu R_\mu^{\;[\nu}\xi^\rho\psi^{\sigma]}=0$, $R_{\mu\nu}$ being the Ricci tensor. For an arbitrary matter distribution in general relativity, this condition simply means that there are no momentum flux or shear stress in the direction of  the meridional planes which are orthogonal to both $\xi^\mu$ and $\psi^\mu$ \cite{Carter:1969zz}. In other words, motion can only be circular around the symmetry axis, and not convective within meridional planes. Under these conditions, the most generic metric can be cast in the following form:
\begin{equation}
\begin{split}
    \mathrm{d}s^2&=g_{tt}(r,\theta)\mathrm{d}t^2+2g_{t\varphi}(r,\theta)\mathrm{d}t\mathrm{d}\varphi+g_{\varphi\varphi}(r,\theta)\mathrm{d}\varphi^2
    \\
    &\quad+g_{rr}(r,\theta)\mathrm{d}r^2+g_{\theta\theta}(r,\theta)\mathrm{d}\theta^2,
    \label{eq:statmet}
    \end{split}
\end{equation}
(see e.g. chapter 7 of Wald's book \cite{Wald:1984rg} or original proofs in Refs.~\cite{Papapetrou:1966zz,Carter:1969zz}). 
It is possible to parametrize $\theta$ such that $0\leq\theta\leq\pi$. We additionally impose a $\mathbb{Z}_2$ symmetry $\theta\to\pi-\theta$, which defines an equator at $\theta=\pi/2$. This symmetry is crucial for the rest of our analysis. Indeed, a particle initially moving in the equatorial direction will remain in the equator plane. For such planar orbits, the calculations are greatly simplified. We will restrict to these orbits for simplicity. Let $u^\mu$ be the tangent vector to some equatorial geodesic, and $\tau$ the affine parameter along this geodesic. The affine parameter can be freely chosen such that
\begin{equation}
g_{\mu\nu}u^\mu u^\nu=-\kappa,
\label{eq:kappa}
\end{equation}
with $\kappa=1$ for timelike geodesics and $\kappa=0$ for null geodesics. In the system of coordinates proposed in Eq.~\eqref{eq:statmet}, Eq.~\eqref{eq:kappa} takes the form
\begin{equation}
-\kappa=g_{tt}\dot{t}^2+2g_{t\varphi}\dot{t}\dot{\varphi}+g_{\varphi\varphi}\dot{\varphi}^2+g_{rr}\dot{r}^2,
\end{equation}
where a dot stands for a derivative with respect to $\tau$. To the two Killing vectors $\xi^\mu=(\partial_t)^\mu$ and $\psi^\mu=(\partial_\varphi)^\mu$ are associated two conserved quantities, $E$ and $L$:
\begin{align}
E&=-g_{\mu\nu}u^\mu \xi^\nu=-g_{tt}\dot{t}-g_{t\varphi}\dot{\varphi},
\label{eq:E}
\\
L&=g_{\mu\nu}u^\mu \psi^\nu=g_{t\varphi}\dot{t}+g_{\varphi\varphi}\dot{\varphi}.
\label{eq:L}
\end{align}
These quantities can be interpreted as the energy and angular momentum per unit mass of a test particle (or $\hbar E$ and $\hbar L$ in the case of a massless particle). We can further define the angular speed of the particle as 
\begin{equation}
\Omega\equiv\dfrac{\mathrm{d}\varphi}{\mathrm{d}t}=-\dfrac{g_{t\varphi}E+g_{tt}L}{g_{\varphi\varphi}E+g_{t\varphi}L}.
\label{eq:Omega}
\end{equation}
Plugging Eqs.~\eqref{eq:E} and \eqref{eq:L} in Eq.~\eqref{eq:kappa}, we obtain
\begin{equation}
g_{rr}\dot{r}^2=\dfrac{g_{\varphi\varphi}E^2+2g_{t\varphi}EL+g_{tt}L^2}{B}-\kappa\equiv -V(r,E,L),
\label{eq:EOM}
\end{equation}
where
\begin{equation}
B(r)\equiv g_{t\varphi}^2-g_{tt}g_{\varphi\varphi}.
\label{eq:B}
\end{equation}
In the domain of outer communications, $B>0$. Indeed, in an asymptotically flat spacetime of the form \eqref{eq:statmet}, the locus of points where $B=0$, if it exists, coincides with the black hole horizon \cite{Carter:1973rla}.
Therefore, outside of a horizon, requiring that the metric has a $(-,+,+,+)$ signature imposes $g_{rr}>0$, $g_{\theta\theta}>0$, $g_{\varphi\varphi}>0$ (the latter condition also following from the definition of axisymmetry). In particular, $g_{rr}>0$ and the test object effectively moves in the 1D potential $V$. It will follow a circular orbit at radius $r$ if 
\begin{align}
    V(r,E,L)&=0,
    \label{eq:V0}
    \\
    V'(r,E,L)&=0,
    \label{eq:dV0}
\end{align}
where a prime denotes a derivative with respect to $r$, keeping other variables fixed. The orbit will be stable if $V''(r,E,L)>0$. Let us now focus our attention on timelike geodesics, with $\kappa=1$. Solving Eqs.~\eqref{eq:V0} and \eqref{eq:dV0} fixes the value of $E$ and $L$. There exists a pair of solutions, corresponding to prograde or retrograde orbits.\footnote{There actually exist two pairs of such solutions, as Eq.~\eqref{eq:EOM} possesses a discrete symmetry $\{E,L\}\to\{-E,-L\}$. We choose the pair which has $\dot{t}>0$, and hence future oriented geodesics.} We will label these as + and $-$ respectively. The solution, if it exists, can be expressed as 
\begin{align}
E_\pm(r)&=-\dfrac{g_{tt}+g_{t\varphi}\Omega_\pm}{\sqrt{\beta_\pm}},
\label{eq:Esol}
\\
L_\pm(r)&=\dfrac{g_{t\varphi}+g_{\varphi\varphi}\Omega_\pm}{\sqrt{\beta_\pm}},
\label{eq:Lsol}
\\
\beta_\pm(r)&\equiv-g_{tt}-2g_{t\varphi}\Omega_\pm-g_{\varphi\varphi}\Omega_\pm^2,
\\
\Omega_\pm(r)&=\dfrac{-g'_{t\varphi}\pm\sqrt{C}}{g'_{\varphi\varphi}},
\label{eq:Omegasol}
\end{align}
where
\begin{equation}
    C(r)\equiv(g'_{t\varphi})^2-g'_{tt}g'_{\varphi\varphi}.
\end{equation}
Clearly, timelike circular orbits only exist when $C>0$ and $\beta_\pm>0$. 
\section*{Stationary and axisymmetric spacetimes}
Let us now establish the important relation that binds the radial variation of the energy to the stability of the orbits. We assume that the particle loses energy adiabatically by gravitational radiation, or by any dissipative process: $\delta E_\pm<0$. \footnote{In vacuum, this assumption is potentially limited by Penrose-like processes, when some particles or waves can carry negative energy. If the object we consider is emitting such negative energy carriers, its energy could be enhanced. However, Penrose process is localized inside ergospheres; if it were to happen in the domain of outer communications, it would most certainly correspond to a fatal instability. Thus, the caveat concerns at most orbits that are enclosed inside the ergosphere. In any case, for a given theory, one can a priori compute a detailed balance of energy gain/loss, and Eq.~\eqref{eq:master} below remains predictive.} At the same time, the radius of the circular orbit will be shifted of $\delta r$. We aim at determining the sign of $\delta r$, to understand under which conditions the particle will spiral inwards. We also assume that the orbit remains circular. We have
\begin{equation}
   0> \delta E_\pm=E'_\pm\delta r,
\end{equation}
hence the sign of $\delta r$ is opposite to the sign of $E'_\pm$. Manipulating  Eqs.~\eqref{eq:Esol}--\eqref{eq:Omegasol}, one can cast $E'_\pm$ into the following, remarkably compact form:
\begin{equation}
    E'_\pm=\dfrac{B}{2\sqrt{C\beta_\pm}}\dfrac{(-g_{tt}')V''}{(\sqrt{C}\pm g_{t\varphi}')},
    \label{eq:master}
\end{equation}
Consider then an asymptotically flat spacetime. Far away from the compact object, $V''>0$, $g'_{tt}<0$, and $\sqrt{C}>|g_{t\varphi}'|$; hence, $E'_\pm>0$ and particles spiral inwards when they lose energy. Is it possible to reach a point where $E'_\pm$ flips sign from such a region? The first factor in the right hand side of Eq.~\eqref{eq:master} is positive definite. The denominator of the second factor can a priori vanish, but given the form of $C$, this can only happen if $g_{tt}'=0$.\footnote{In whole generality, $g_{\varphi\varphi}'$ could also vanish, but this in general indicates a breakdown of the system of coordinates, see e.g. Ref.~\cite{Wald:1984rg}.} There exist therefore only two possibilities for $E'_\pm$ to change sign:

(i) The first is that $V''$ changes sign, meaning the orbits transition from stable to unstable. Adiabatic evolution will stop whenever this happens, and be followed by a plunge. Indeed, if the particle started with a unit energy far away, its energy only decreased since, and it can therefore not be expelled towards spatial infinity. This is what occurs typically at the ISCO of standard general relativity black holes.

(ii) The second possibility is that $g'_{tt}$ vanishes. In this case, as mentioned above, one needs to take into account the denominator of the second factor in Eq.~\eqref{eq:master}. If $g_{t\varphi}'>0$ at the point where $g'_{tt}=0$, only the retrograde orbit will undergo a change of sign in $E'$. Nothing particular happens to particles in prograde orbits at this location. If $g_{t\varphi}'<0$, the result is simply exchanged between prograde and retrograde orbits. Let us focus on the case $g_{t\varphi}'>0$ for concreteness.

Remarkably, when $g'_{tt}=0$, the retrograde rotation speed $\Omega_-$ vanishes as well, as can be seen from Eq.~\eqref{eq:Omegasol}. This orbit, when it exists, is called a static ring. The possibility of having static rings was considered in \cite{Collodel:2017end} and further studies \cite{Teodoro:2020kok,Teodoro:2021ezj,Collodel:2021jwi}. Typically, the metric potential $g_{tt}$ possesses maxima when the matter distribution is off-centered, like in toroidal boson stars configurations \cite{Kleihaus:2005me}, or in rapidly rotating neutron stars with differential rotation (see e.g. \cite{Komatsu:1989zz}). Neutron stars have not been investigated in this respect, but the relevant configurations usually correspond to short-lived merger remnants. Therefore, a detailed analysis of their geodetic properties does not seem opportune. In the case of boson stars on the other hand, an explicit example where $g_{tt}$ possess a minimum in a region where stable timelike circular orbits exist is provided in Ref.~\cite{Collodel:2021jwi}. Such orbits present outstanding features; they correspond to points of accumulation for infalling objects. An object placed in a circular orbit slightly below the static ring would \textit{outspiral} up to the static ring. Particles approaching this orbit would exhibit a very specific gravitational wave phase, with a frequency progressively vanishing, hence getting out of the sensitivity range of any detector. We will see in the next paragraph that these features are in fact not specific to rotating spacetimes, and persist in spherically symmetric geometries, under the same condition, i.e. $g_{tt}'=0$.
 
Note that nothing prevents $g_{tt}$ from having several minima and $g_{t\varphi}'$ from changing sign along the radial direction. Thus, a given spacetime can have several static rings, either for prograde or retrograde orbits.

Of course, the sequence of circular timelike orbits can also be interrupted abruptly if $C$, $B$ or $\beta_\pm$ vanishes. The particle will then plunge deeper towards the center of the geometry, potentially reaching another region where circular timelike orbits exist, with lower energy. In fact, the particle may even acquire a negative energy. Examples of stable timelike circular orbits with negative energies were exhibited in \cite{Ansorg} in the context of dust disks, or more recently in \cite{Collodel:2021jwi} for black holes with bosonic hair. However, a particle may only acquire negative energy on its orbit inside an ergosphere. Indeed, when $E_\pm=0$,
\begin{equation}
\beta_\pm=\dfrac{g_{tt}}{g_{t\varphi}^2}B,
\end{equation}
hence $g_{tt}$ must be positive (note that $\beta_\pm$ and $B$ must be positive, see Eq.~\eqref{eq:Esol} and corresponding discussion). This is reassuring: particles with negative energies emerge rather as an artifact due to the spacelike character of the $\xi^\mu$ direction than as a source of instability which could destroy the geometry.

In passing, we remark that the energy of a particle on a static ring is 
\begin{equation}
    E_\mathrm{static}=\sqrt{-g_{tt}},
    \label{eq:Estatic}
\end{equation}
i.e. simply the local redshift factor. Equation \eqref{eq:Estatic} also confirms that a static ring cannot be located inside an ergosphere, as should be expected from the very nature of an ergosphere.

\section*{Spherically symmetric and static spacetimes}
\label{sec:spheric}
The discussion that follows Eq.~\eqref{eq:master} is even simpler when considering spherically symmetric and static spacetimes. Indeed, we can make the choices $g_{t\varphi}=0$ and $g_{\varphi\varphi}=r^2$ in Eq.~\eqref{eq:statmet}. Considering equatorial geodesics is obviously not a restriction in this case. Equation \eqref{eq:master} simplifies to
\begin{equation}
    E'=\dfrac{r}{4}V''E,
    \label{eq:masterspheric}
\end{equation}
while $E>0$ always. We dropped the $\pm$ index, since prograde and retrograde orbits have the same energy in spherical symmetry.
Hence, Eq.~\eqref{eq:masterspheric} unambiguously states that particles losing energy have to spiral inwards, either until they reach the origin of the system of coordinates, or they reach an unstable orbit and subsequently plunge closer to the center. 

Note that $g_{tt}'$ has disappeared from Eq.~\eqref{eq:masterspheric}. In spherical symmetry, the condition $C>0$, required for circular orbits to exist, becomes $g_{tt}'<0$. It is therefore legitimate to wonder what happens for a putative metric where $g_{tt}'$ vanishes at some radius. For instance, one can view thin-shell gravastars as a limit case of a metric with $g_{tt}'=0$ on the shell \cite{Visser:2003ge}. Computing $V''$ at an extremum of $g_{tt}$, we obtain
\begin{equation}
    V''=\dfrac{g_{tt}''}{g_{tt}}.
    \label{eq:stabmin}
\end{equation}
Keeping in mind that $g_{tt}<0$ outside of a horizon, minima of $g_{tt}$ always correspond to unstable orbits, while maxima are always stable orbits. A typical example where $g_{tt}$ has a minimum is the Schwarzschild-de Sitter metric. Indeed, in this case, the minimum of $g_{tt}$ is always located at a larger radius than the outermost stable circular orbit, and hence it cannot correspond to a stable orbit \cite{1979AuJPh..32..293H}. The situation is more interesting if $g_{tt}$ possesses a maximum. The corresponding orbit will be stable as a consequence of Eq.~\eqref{eq:stabmin}, and it will have $\Omega_\pm=0$, $E=\sqrt{-g_{tt}}$. Thus, the concept of static ring is not specific to rotating spacetimes. A particle spiraling down from infinity will also freeze on a ring (or rather a sphere) where $g_{tt}'=0$ in spherical symmetry. Contrary to what happens in stationary spacetimes, this sphere also corresponds to the limit of existence of circular orbits. Hence, one might be worried that perturbing the orbit inwards will cause the particle to fall. This is not the case. Indeed, let us call $r_0$ the radius of the point where $g_{tt}'=0$. Perturbing the position of the particle according to $r_0\to r_0+\delta r(\tau)$, we get at first order
\begin{equation}
    \ddot{\delta r}+\omega^2\delta r=0,
\end{equation}
where
\begin{equation}
    \omega^2=\left.\dfrac{g_{tt}''}{g_{rr}g_{tt}}\right|_{r_0}>0.
\end{equation}
Hence, the particle will oscillate around $r_0$, while possibly revolving around the central object. It will penetrate the region where circular orbits are forbidden, but not on a circular orbit.
\section*{Unbound ISCOs}
\begin{figure*}[ht]
\begin{center}
	\includegraphics[width=0.5\textwidth]{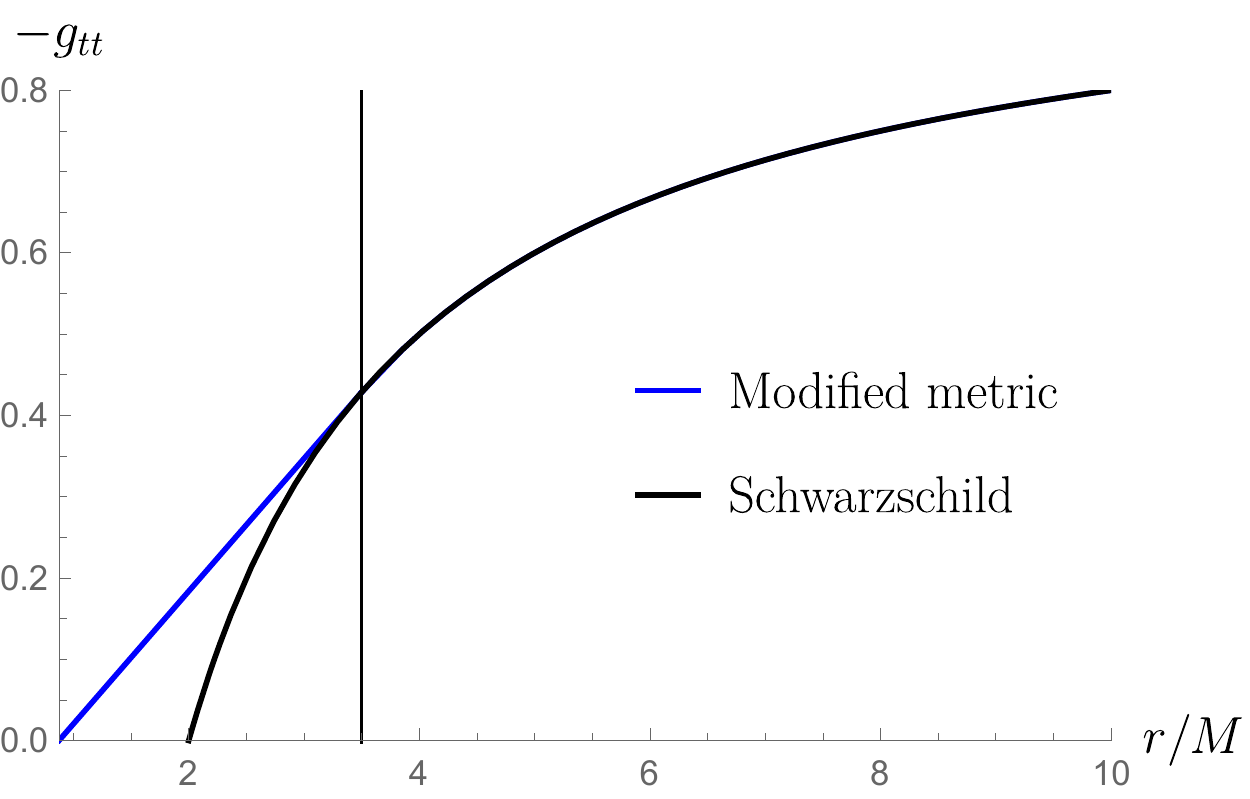}%
	\includegraphics[width=0.5\textwidth]{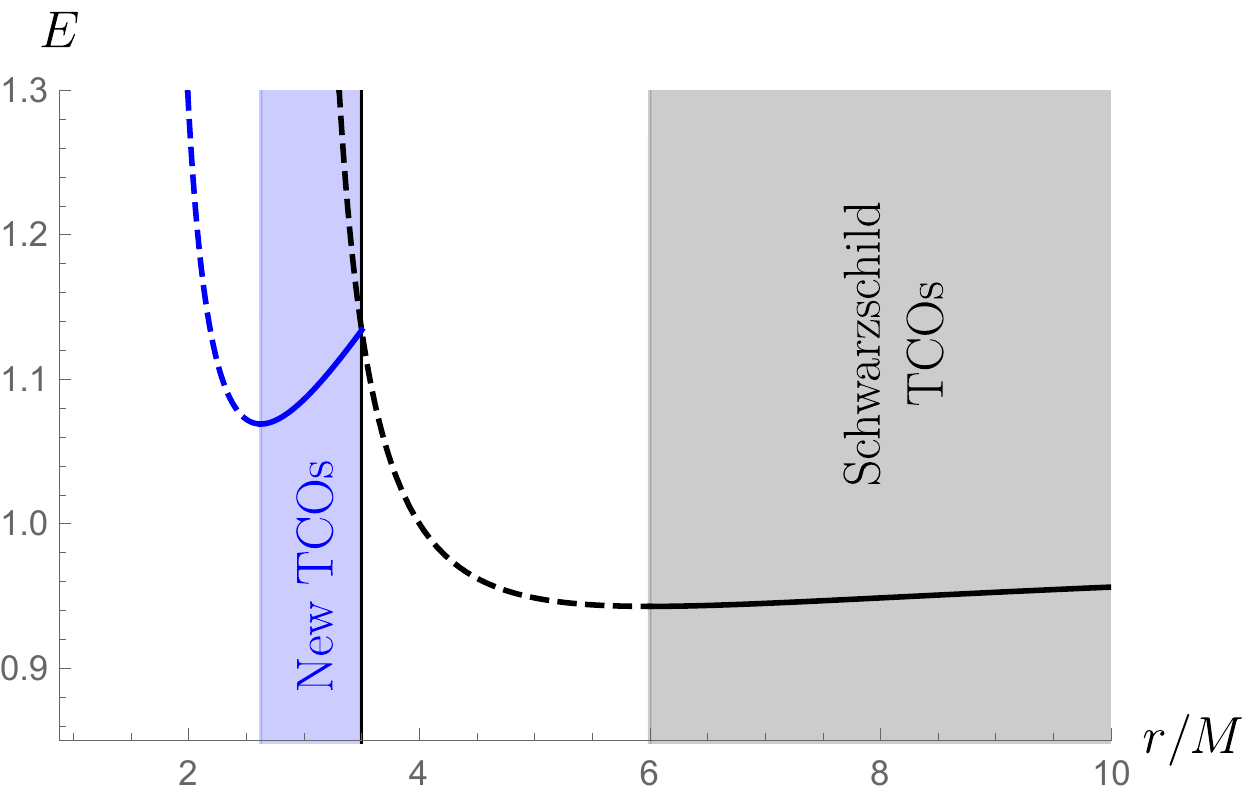}%
	\end{center}
	\caption{Comparison between Schwarzschild metric, in black, and the modified metric of Eqs.~\eqref{eq:gttS}-\eqref{eq:gttmod} with $\epsilon=3.5$, in blue. The left panel shows the metric profile, while the right panel shows the energy profile. The modified metric differs from Schwarschild's only for $r<\epsilon M$; this radius is indicated by a vertical black line. The geodetic structure is very different. Besides the region where stable timelike circular orbits exist in Schwarzschild spacetime ($r>6M$, gray region in the right panel), additional stable orbits exist for $r$ slightly smaller than $\epsilon M$ (region shaded in blue). In the right panel, plain lines correspond to stable orbits, while dashed lines indicate unstable orbits. The stability can be inferred directly from the slope of $E$, as per Eq.~\eqref{eq:masterspheric}. The additional region of stable orbits at $r<\epsilon M$ generates an unbound ISCO, with $E_\text{ISCO}>1$. In passing, the modified metric possesses a light ring at $r=\epsilon(4-\epsilon)M$.}
	\label{fig:unbISCO}
\end{figure*}

Next, we aim at addressing the following question: can a compact object possess an unbound ISCO, i.e. be such that particles which spiraled down to this orbit have enough energy to be ejected at spatial infinity? 
Obviously, if the spacetime is asymptotically flat and if it is possible to reach the ISCO spiraling down through a sequence of stable circular orbits, particles will be bound there. In other words, starting from $E=1$ far away and imposing $\delta E<0$, one necessarily ends up with $E_\text{ISCO}<1$. If there exist unbound ISCOs, they cannot be standard ISCOs in the language of Ref.~\cite{Delgado:2021jxd}. However, there can exist several, disconnected domains of stable circular orbits. We will now construct explicitly a toy metric which displays an unbound ISCO. Although it is not a solution of general relativity equations, we will see that it is extremely close to the Schwarzschild metric and respects the (geometric) energy conditions. We will glue smoothly the Schwarzschild metric with a linear profile at a radius $r=\epsilon M$, and adjust the value of $\epsilon$ to produce new timelike circular orbits. Namely, we choose
\begin{align}
    -g_{tt}&=1-\dfrac{2M}{r}~\text{if}~r\geq\epsilon M,
    \label{eq:gttS}
    \\
       - g_{tt}&=a+br~\text{if}~r<\epsilon M
            \label{eq:gttmod}
\end{align}
and $g_{rr}=-1/g_{tt}$. The coefficients $a$ and $b$ are fixed by requiring that the metric and its derivative are continuous at $r=\epsilon M$:
\begin{equation}
    a=\dfrac{\epsilon-4}{\epsilon},\qquad b=\dfrac{2}{\epsilon^2M}.
\end{equation}
This modified metric displays a horizon at $r=\epsilon(4-\epsilon)/2$. We fill focus on choices of $\epsilon$ smaller than 4, so that the areal radius of 2-spheres does not become zero before reaching the horizon (we will also ignore what happens inside the horizon for our analysis). As a consequence, the metric is identical to Schwarzschild at least for $r\geq4M$. Hence it displays a standard ISCO at $r=6M$. It is very easy to show that new stable circular orbits (characterized by Eqs.~\eqref{eq:V0}-\eqref{eq:dV0} and $V''>0$) appear in the inner region \eqref{eq:gttmod} if and only if $10/3<\epsilon<4$. These new orbits exist over the range
\begin{equation}
 \dfrac32\epsilon(4-\epsilon)M<r<\epsilon M.
\end{equation}
We can therefore compute the energy at the new, absolute ISCO. It is given by
\begin{equation}
    E_\text{ISCO}=2\sqrt{\dfrac{2(4-\epsilon)}{\epsilon}}.
\end{equation}
This energy is greater than 1 for $\epsilon\leq32/9$. Therefore the metric \eqref{eq:gttS}-\eqref{eq:gttmod} exhibits an unbound absolute ISCO if and only if
\begin{equation}
    \dfrac{10}{3}\simeq3.33<\epsilon<\dfrac{32}{9}\simeq3.56.
\end{equation}
This is illustrated in Fig.~\ref{fig:unbISCO}.
Beyond the fact that it reproduces the Schwarzschild solution at $r\geq4M$, a very important feature of the metric \eqref{eq:gttS}-\eqref{eq:gttmod} is that it respects the weak and strong energy conditions, in their geometric form. These conditions can be formulated respectively as $G_{\mu\nu}t^\mu t^\nu\geq0$ and $R_{\mu\nu}t^\mu t^\nu\geq0$ for any timelike vector $t^\mu$. These conditions are expected to hold for any reasonable type of matter. In particular, in general relativity, the weak energy condition asserts that the energy density of matter, as measured by an observer with 4-velocity $t^\mu$, is positive. In spherical symmetry, the weak energy condition translates as:
\begin{align}
    0&\leq (g_{rr}-1) g_{rr} + r g_{rr}',
    \label{eq:WEC1}
    \\
0&\leq -g_{tt} g_{rr}' - g_{rr} g_{tt}',
\label{eq:WEC2}
\\
\begin{split}
0&\leq -\left\{-4 g_{rr}^2 g_{tt}^2 + 
    r g_{tt} g_{rr}'(-2 g_{tt} + r g_{tt}')\right. \\
    &\quad\left.+ 
    g_{rr} [4 g_{tt}^2 + r^2 g_{tt}'^2 - 
       2 r g_{tt}(g_{tt}' + 
          r g_{tt}'')]\right\}.
          \label{eq:WEC3}
\end{split}
\end{align}
The strong energy condition amounts to Eqs.~\eqref{eq:WEC2}-\eqref{eq:WEC3} together with 
\begin{equation}
    0\leq2 r g_{rr} g_{tt} g_{tt}''-g_{tt}'\left[r g_{tt} g_{rr}'+g_{rr} \left(r g_{tt}'-4 g_{tt}\right)\right].
    \label{eq:SEC}
\end{equation}
The right hand side of all above equations vanishes at $r\geq\epsilon M$ since there the solution is locally identical to the Schwarzschild metric. The right hand side of Eq.~\eqref{eq:WEC2} still vanishes below $\epsilon M$. It is straightforward to check that the right hand sides of Eqs.~\eqref{eq:WEC1}, \eqref{eq:WEC3} and \eqref{eq:SEC} are strictly positive for the metric \eqref{eq:gttmod} down to the horizon. Hence, we can a priori construct the geometry \eqref{eq:gttS}-\eqref{eq:gttmod} with a physically acceptable type of matter.

Therefore, a very moderate deformation of the Schwarzschild metric is sufficient to equip the spacetime with an unbound ISCO. Such an ISCO cannot be reached from spatial infinity through an adiabatic sequence of timelike orbits. Rather, it would have to be populated by debris of infalling objects. This could result for instance from the collision of objects with highly eccentric orbits. Objects that end up on this unbound ISCO could then escape the gravity of the central object at the end of their inspiral, against Newtonian intuition.

\section*{Conclusions}

It is remarkable that it is possible to derive these results for an arbitrary theory of gravity which describes spacetime through a metric. The fate of an observer in circular motion is determined by two ingredients: i) the weak equivalence principle i) the fact that the observer will lose energy in a dissipative process. Given these two conditions, the slope of the metric potential $g_{tt}$ will determine the evolution of the trajectory. Wherever $g_{tt}'<0$, circular orbits, if they exist, will tend to shrink. Orbits may grow when $g_{tt}'>0$, although such spacetimes are not common. In this case, maxima of $g_{tt}$ become accumulation points for massive particles. This is true for whatever field content or intricate equations describing the dynamics of the metric. 

The proof importantly relies on the assumption that the test-body is only losing energy. Hence, it does not cover cases where energy is injected in the system from an external or internal source. A typical internal source, present already in Newtonian gravity, is the spin of the orbiting bodies. The tidal acceleration phenomenon can cause the central body to spin down while the gravitational energy of the satellite increases. Because of this phenomenon, the Moon is moving away from the Earth \cite{2016CeMDA.126...89W}. External sources of energy can become important if their frequency resonates with the characteristic frequencies of the orbiting object. These sources can be for instance a gravitational wave background \cite{Blas:2021mpc} or potentially a distant third object \cite{Bonga:2019ycj}.
However, we emphasize that, even in these cases, the knowledge of the detailed energy balance together with our analysis allows one to predict the evolution of the orbit.

We also proposed a toy metric that features an unbound ISCO, hence allowing non-standard behaviors for objects that terminate their inspiral there. This metric describes a black hole identical to the Schwarzschild one except in a narrow region close to the horizon. It respects the weak and strong energy conditions.

\acknowledgments{We thank Panagiotis Iosif and Nikolaos Stergioulas for helpful discussions on the properties of rapidly rotating neutron stars. This project has received funding from the European Union's Horizon 2020 research and innovation program under the Marie Sklodowska-Curie grant agreement No 101007855. 
We thank FCT for financial support through Project~No.~UIDB/00099/2020.
We acknowledge financial support provided by FCT/Portugal through grants PTDC/MAT-APL/30043/2017 and PTDC/FIS-AST/7002/2020.
We also acknowledge  networking support by the GWverse COST Action
CA16104, ``Black holes, gravitational waves and fundamental physics.''
}

\bibliographystyle{unsrt}
\bibliography{biblio}
\end{document}